\documentclass[a4paper,11pt]{article}
\pdfoutput=1 

\usepackage{jheppub} 

\usepackage[T1]{fontenc} 
\usepackage{graphbox}
\usepackage{lineno}
\usepackage{placeins}

\newcommand{\ie}{{\sl i.e.}~}
\newcommand{\eg}{{\sl e.g.}~}
\newcommand{\pt}{\ensuremath{p_{\rm T}}}
\newcommand{\MadGraph}{\textsc{MadGraph5}}
\newcommand{\Pythia}{\textsc{Pythia8}}
\newcommand{\Herwig}{\textsc{Herwig7}}
\newcommand{\Delphes}{\textsc{Delphes3}}
\newcommand{\PowhegBox}{\textsc{POWHEG-BOX}}
\newcommand{\McAtNLO}{\textsc{MC@NLO}}
\newcommand{\GEANT}{\textsc{GEANT4}}
\newcommand{\Sherpa}{\textsc{Sherpa}}

\title{\boldmath DijetGAN: A Generative-Adversarial Network Approach for the Simulation of QCD Dijet Events at the LHC}

\author[1]{Riccardo Di Sipio,}
\author[2]{Michele Faucci Giannelli,}
\author[1]{Sana Ketabchi Haghighat,}
\author[2]{Serena Palazzo.}

\affiliation[1]{University of Toronto, Canada}
\affiliation[2]{University of Edinburgh, UK}

\emailAdd{riccardo.disipio@utoronto.ca}
\emailAdd{michele.faucci.giannelli@ed.ac.uk}
\emailAdd{sana.ketabchihaghighat@mail.utoronto.ca}
\emailAdd{serena.palazzo@ed.ac.uk}

\abstract{A Generative-Adversarial Network (GAN) based on convolutional neural networks is used to simulate the production of pairs of jets at the LHC. The GAN is trained on events generated using \textsc{MadGraph5}, \textsc{Pythia8}, and \textsc{Delphes3} fast detector simulation. We demonstrate that a number of kinematic distributions both at Monte Carlo truth level and after the detector simulation can be reproduced by the generator network with a very good level of agreement.
\paragraph{}
The code can be checked out or forked from the publicly accessible online repository \url{https://gitlab.cern.ch/disipio/DiJetGAN}.}

\begin{document} 
\maketitle
\flushbottom

\section{Introduction}
\label{sec:intro}
In the forthcoming years, experiments at the Large Hadron Collider (LHC) are expected to cope with a deluge of data. At the same time, the strategy to produce reliable and statistically large samples of simulated  proton--proton ($pp$) inelastic collisions will be facing both technological limitations and new opportunities.

In the context of Machine Learning (ML) and specifically Deep Neural Networks (DNN), an unsupervised learning technique called Generative-Adversarial Networks (GAN) has been proposed recently \cite{GAN} to create a sample based on a set of unlabelled training examples. In practice, GANs have been widely used to generate photorealistic portraits \cite{GAN:images}, music \cite{GAN:music}, and recently also to simulate the response of a calorimeter to the passage of particles \cite{GAN1,GAN2,GAN3,GAN4,GAN5,GAN6,GAN7}. Other deep learning methodologies such as variational autoencoders (VAE) \cite{VAE} have been applied to reproduce several kinematic distributions learned from examples taken from Monte Carlo simulations. The same study also reports a less satisfactory performance when a GAN based on fully-connected deep networks was used \cite{BetaVAE}. In another work \cite{GAN8}, authors employed a similar architecture, but also taking into account the symmetries of the process under consideration. The events used for the training are pre-processed so that the azimuthal angle of the leading charged lepton is always zero, which leads to a reduction of the degrees of freedom of the problem. Despite the fact that a substantial improvement is observed in terms of agreement with the testing sample, the distribution of the azimuthal angle is not uniform, indicating that the network failed to learn the left-right symmetry. In this work, we successfully trained a GAN to reproduce kinematic distributions, improving on such previous attempts, thanks to a careful consideration of a larger number of implicit symmetries of the physics process under study, and the employment of deep convolutional layers \cite{DCGAN} following the example of \cite{JetTagging2}.
\paragraph{}
In a GAN, a generative network $G$ transforms a vector of random numbers (input noise) $z\sim p_z$ into a sample carrying some physical meaning, which in this case are the four-momenta of the two jets. In practical applications, $p_z$ is usually a $N$-dimensional uniform distribution in the range $[0,1]^N$.  Subsequently, a discriminative network $D$ estimates the probability that a given sample comes either from the training data or the generator. The two samples are distributed with probability density functions ({\sl pdf}) $p_{data}$ and $p_{fake}$ respectively, with $p_{data}$ fixed and usually estimated using a Monte Carlo method. The Nash equilibrium (min-max game) is reached when $D$ is unable to distinguish fake examples from real data, hence the generator has been trained to be a good approximator of the data {\sl pdf}, \ie $p_{fake}\sim p_{data}$. The procedure is equivalent to the minimization of the loss function $L(G,D)$ :

\begin{eqnarray}
    \min_{G} \max_{D} L(G,D) & = & \mathbb{E}_{x\sim p_{data}(x)}\left [ \log D(x)\right ] + \mathbb{E}_{z\sim p_{z}(z)}\left [ \log \left ( 1 - D(G(z)\right ) \right ] \\
    & = & \mathbb{E}_{x\sim p_{data}(x)}\left [ \log D(x)\right ] + \mathbb{E}_{x\sim p_{fake}}(x)\left [ \log \left ( 1 - D(x)\right ) \right ] \\
\end{eqnarray}

For a fixed $G$, the optimal discriminator $D^\star_G$ is given by a monotonic function of:

\begin{equation}
    D^\star_G(x) = \frac{p_{data}(x)}{p_{data}(x) + p_{fake}(x)}
\end{equation}

Under this condition, the min-max game can be reformulated as the minimization of the Jensen-Shannon divergence $JS$, which can be expressed in terms of the Kullback-Leibler divergence $KL$, \ie a measure of how one probability distribution is different from a second:

\begin{eqnarray}
    KL(p_{data}|| p_{fake}) & = &  \mathbb{E}\left [ \log p_{data} - \log p_{fake} \right ] \\
    JS(p_{data}|| p_{fake}) & = & \frac{1}{2}KL\left ( p_{data}||\frac{p_{data}+p_{fake}}{2}\right ) + \frac{1}{2}KL\left ( p_{data}||\frac{p_{data}+p_{fake}}{2}\right ) \\
    \min_G L(G,D^\star) & = & 2 \min_G JS(p_{fake}||p_{data}) -2\log 2
\end{eqnarray}

The optimal generator $G^\star$ is such that $p_{fake} = p_{data}$, hence the minimum of the loss function is achieved, \ie $D^\star_{G^\star}(x)= 1/2$, $JS(p_{fake}||p_{data})=0$ and $L(G^\star,D^\star)=-2\log2$.

\paragraph{}
In this article, we introduce the architecture of a GAN that aims to simulate the production of pairs of particles in $pp$ interactions at the LHC. In particular, this approach is applied to dijet production, which is a ubiquitous source of background to Standard Model precision measurements \cite{SM1, SM2, SM3} and searches for physics beyond the Standard Model \cite{BSM1,BSM2,BSM3,BSM4}. 

\paragraph{}
Currently, both the ATLAS \cite{ATLAS} and CMS \cite{CMS} experiments of the LHC at CERN deploy Monte Carlo (MC) event generators such as \MadGraph~\cite{MadGraph}, \PowhegBox~\cite{POWHEG-BOX}, \McAtNLO~\cite{MCatNLO}, \textsc{Pythia8} \cite{Pythia8}, \Herwig~\cite{Herwig7} and \Sherpa \cite{Sherpa} to simulate the hard-scattering (HS) and the parton-shower (PS) processes, and a \GEANT~\cite{Geant4} simulation of the actual detector for the response of the experimental apparatus. Best estimates suggest that the simulation of a single event takes already several minutes \cite{hsf-mc}, with O($10^9$) events to be generated for each simulation campaign, leading to a huge computational footprint both in terms of CPU usage and disk space. The situation is particularly critical for processes that require the matching of fixed-order next-to-leading matrix element calculations to parton shower ("multijet merging"), as implemented for example in the \textsc{FxFx} \cite{fxfx}, \textsc{MePS\@NLO} \cite{mepsnlo} and \textsc{UNLoPS} \cite{unlops} methods. Currently, such advanced MC generators cannot be used as the standard generators by the LHC collaborations due to the large time required to generate the billions of events normally required. Hence the experiments need to rely on simpler, but less accurate, generators. Providing a solution to extend the simulated events to the requirements of the LHC experiments will significantly enhance a wide range of measurements.
Similarly, detailed simulation based on \GEANT{} will not be an affordable solution due to the large time required to simulate an event \cite{hsf-mc}. Both experiments are already using fast simulation and are developing new tools exploiting ML and other advanced statistical techniques. We will demonstrate that the same GAN used for reproducing the generator output can be also used to reproduce a simulation of a detector response with a significant time gain with respect to full simulation.   
\paragraph{}
The code is publicly accessible on the online repository  \url{https://gitlab.cern.ch/disipio/DiJetGAN}.

\section{Physics of QCD dijet events}
At hadron colliders such as the LHC, the most abundant kind of interaction between the two colliding protons is the scattering between quarks and gluons  (collectively referred to as {\sl partons}). According to calculations based on the SM, these parton-scattering processes via strong interactions described by quantum chromodynamics (QCD) result in the overwhelming majority of cases in two outgoing partons which carry a net color charge and evolve from high to low virtuality producing parton showers, which eventually hadronize into collimated highly-energetic clusters of particles called {\sl jets}. Hence, 2$\rightarrow$2 parton scattering processes with a pair of jets in the final state are called {\sl dijet events}. The relationship between the clusters and the original partons is revealed by the execution of a clustering algorithm \cite{jetography}. One can think of a jet approximately as a cone of radius $R$ whose axis correspond to the direction of flight of the initial parton. The size of the radius can be controlled by setting a distance parameter in the clustering algorithm. 
\paragraph{}
 For most analyses, the most relevant jets are produced with a large transverse momentum (\pt) and large angle with respect to the incoming partons. The jet mass, defined as the norm of the four-momentum sum of constituents inside a jet, is only loosely related to the mass of the originating parton, and comes mostly from the dynamics of strong interactions. Programs such as \Pythia~\cite{Pythia8} and \Herwig~\cite{Herwig7} implement such calculations with beyond the leading-logarithm (LL) accuracy in what are called a Parton Shower algorithms. The jet mass also plays a key role in the identification of Lorentz-boosted hadronically decaying massive particles such as top quarks \cite{SM1,BSM3}, vector ($W$ and $Z$) and Higgs bosons \cite{BSM5}.
 Finally, QCD predicts a characteristically smooth and monotonically decreasing distribution for the dijet invariant mass ($m_{jj}$), and small production angle. Instead, many theories beyond the SM predict the presence of additional massive particles decaying to dijet pairs, whose hypothetical presence would distort both the dijet invariant mass and production angle distributions in model-dependent ways \cite{BSM1,BSM2}. It is therefore extremely important to be able to reproduce these distributions in a simulation, with particular emphasis on the region of the phase space with $m_{jj}>$~1 TeV, where signs of physics beyond the Standard Model may become evident.
 \paragraph{}
 In the following sections, the agreement between MC calculations and the output of the GAN is evaluated by comparing the individual jets' and dijet system's transverse momentum, pseudo-rapidity\footnote{Pseudorapidity is a commonly spatial coordinate describing the angle of a particle relative to the beam axis defined as $\eta = \frac{1}{2} \ln\frac{E+p_L}{E-p_L}$, where $E$ is the energy and $p_L$ is the longitudinal component of the momentum. It is related to the other components of the momentum via the relationship $|p|=p_T\cosh\eta$.} ($\eta$) and mass distributions. The $\chi^2$ between the MC and the GAN distributions is used as figure of merit.

\section{Monte Carlo Sample}
A sample of 10 million dijet events has been generated using \MadGraph~ and \Pythia, corresponding to an integrated luminosity of about 0.5 fb$^{-1}$. The response of the detector was simulated by a \Delphes~\cite{Delphes} fast simulation, using settings that resemble the ATLAS detector. An average of 25 additional soft-QCD $pp$ collisions (pile-up) were overlaid to reproduce more realistic data-taking conditions. 

Electrons, muons, jets and missing transverse energy are reconstructed by \textsc{Delphes3} algorithms before and after the detector simulation. These two levels of reconstruction are referred to in the following sections as particle- and reco-level respectively. At particle-level, only stable final-state particles, \ie particles that are not decayed further by the generator, and unstable particles\footnote{Particles with a mean lifetime $\tau$ > 300 ps} that are to be decayed later by the detector simulation, are considered. Jets were reconstructed using the anti-$k_T$ algorithm \cite{antikt} as implemented in FastJet \cite{fastjet}, with a distance parameter $R=1.0$.

To increase the number of events with both jets with $p_T > 250$ GeV, a cut on the scalar sum of the transverse momenta of the outgoing partons $H_T > 500$ GeV was applied. Approximately 7,5 million events passed this selection at particle level, and about 4 million at reco level. These events were used to train the network in the subsequent steps.

\section{Network Architecture}\label{sec:architecture}

\begin{figure}[bthp]
\centering
\includegraphics[align=t,scale=0.23]{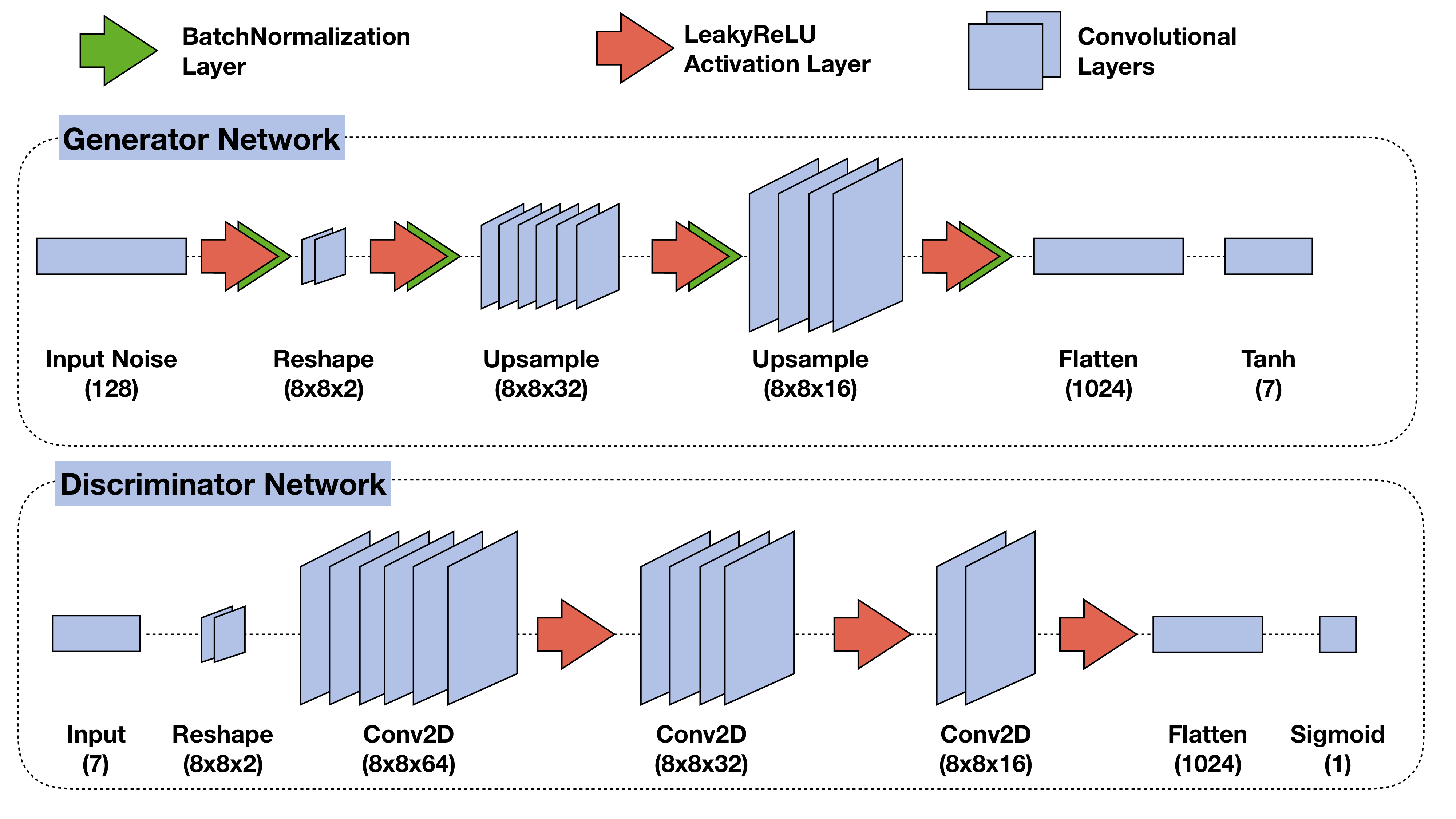}
\caption{\small{Network architecture: generator (top), discriminator (bottom). The GAN is composed by connecting the output of the generator to the input of the discriminator.}}
\label{fig:network}
\end{figure}

The overall architecture of the network, summarized in Fig. \ref{fig:network}, is composed of two main blocks: a generator (G) and a discriminator (D), based on convolutional layers. All layers have {\sl LeakyReLU} activation functions \cite{LeakyReLU} except the last layers that have either {\sl tanh} or {\sl sigmoid} for the generator and the discriminator respectively. The generator transforms a vector of 128 random numbers drawn from a uniform distribution in the $[0,1]$ range into a vector of 7 elements representing the $p_T$, $\eta$ and mass of the leading jet, and the $p_T$, $\eta$, $\phi$ and mass of the second-leading jet.
The network is implemented and trained using \textsc{Keras} v2.2.4 \cite{Keras} with \textsc{Tensorflow} v1.12 \cite{tensorflow} back-end. Input features are scaled in the range $[-1,1]$ and pre- and post-processed using the \textsc{scikit-learn} \cite{scikit-learn} and \textsc{Pandas} \cite{pandas} libraries. The loss function of the generator is mean squared error, while that of the discriminator is the binary cross-entropy. The optimizer is in both cases Adam \cite{adam} with learning rate $lr=$ 10$^{-5}$, $\beta_1$ = 0.5 and $\beta_2$ = 0.9. 
The parameters described above are those that provide the best results among many values and configurations tested. Having reached a satisfactory performance, no further parameter optimisation was carried out.

\section{Training}\label{sec:training}

For the purpose of the training, all MC events were rotated so that the azimuthal $\phi$ angle of the leading jet is always zero. A significant performance improvement was achieved by exploiting the intrinsic  $\phi$ symmetry in di-jet events; the $\phi$ of the leading jet is set to zero while the $\phi$ of the other jet is set to the absolute value of the difference in $\phi$ between the two generated jets. This transformation is reversed when events are generated. In order to further deploy the symmetries of dijet kinematics, every event is used twice: first in its original configuration, and then with the sign of the pseudorapidity of each jet reversed ($\eta$-flip). During event generation, the $\eta$ of the jets is randomly flipped to remove any nonphysical effects that could be introduced by the GAN.

\begin{figure}[htbp]
\centering
\includegraphics[scale=0.20]{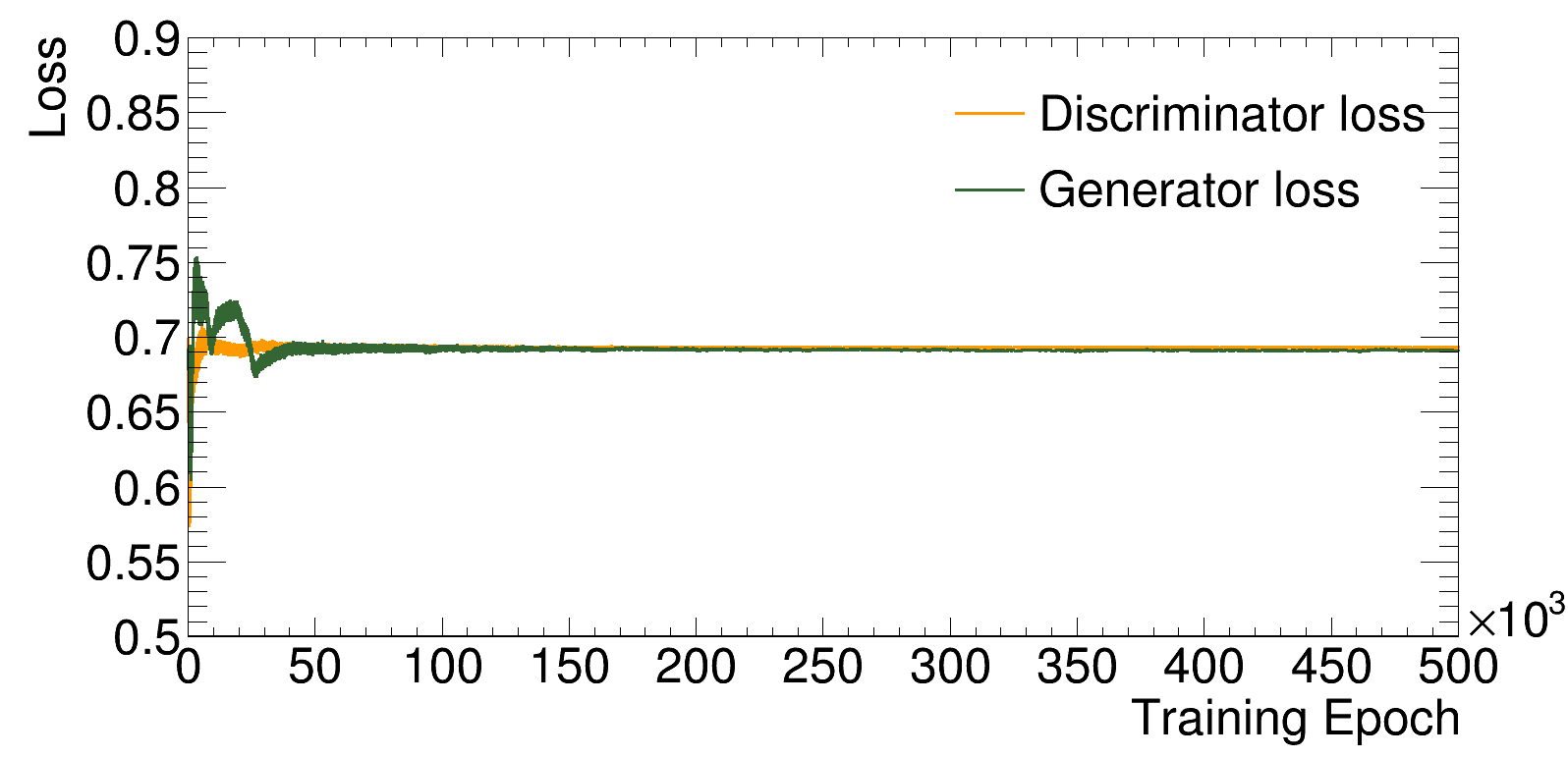}
\caption{\small{Evolution of the discriminator and generator loss at particle level as a function of the training epoch.}}
\label{fig:training:ptcl}
\end{figure}

\begin{figure}[htbp]
\centering
\includegraphics[scale=0.20]{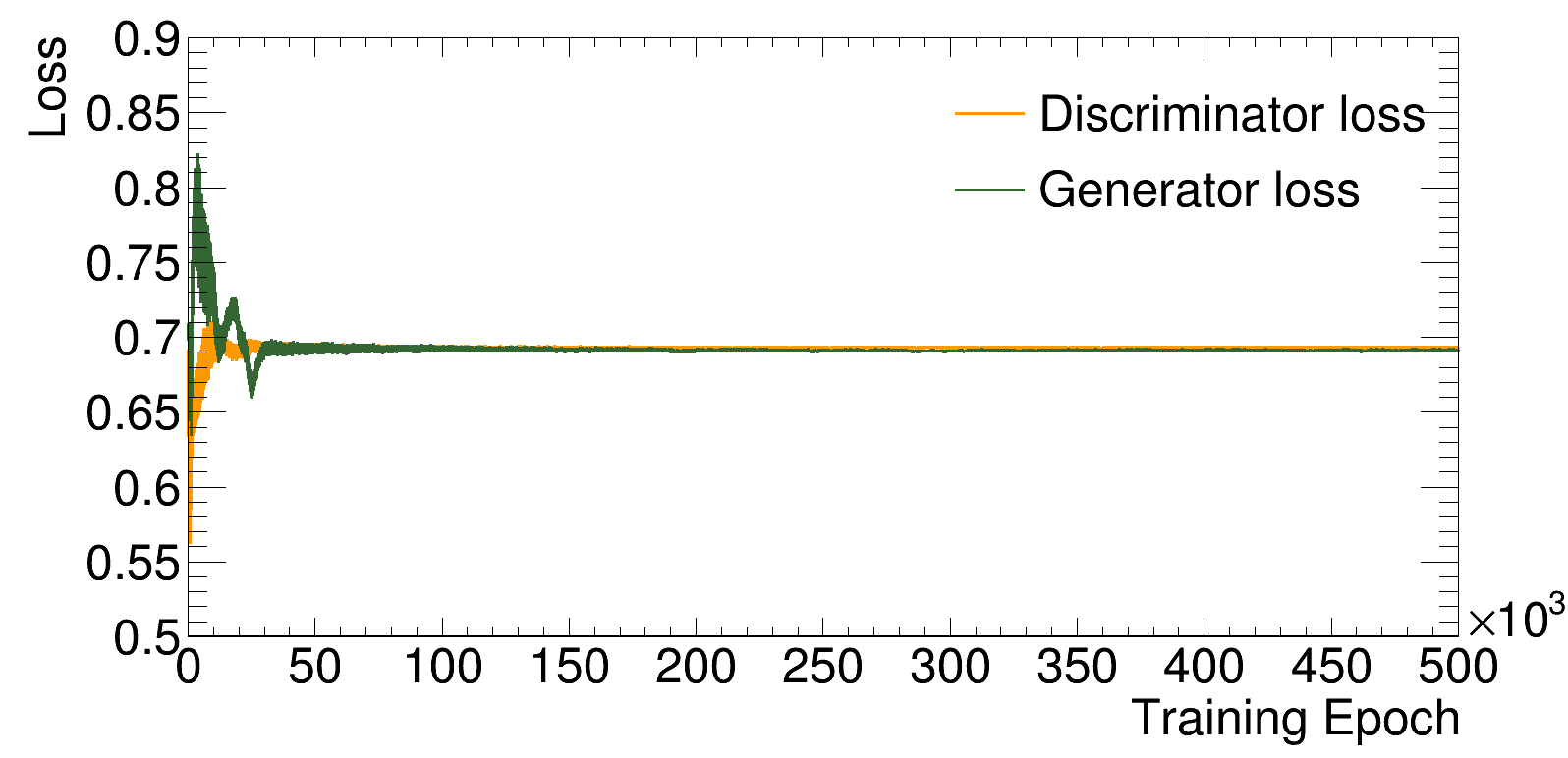}
\caption{\small{Evolution of the discriminator and generator loss at reco level as a function of the training epoch.}}
\label{fig:training:reco}
\end{figure}

\begin{figure}[htbp]
\centering
\includegraphics[scale=0.60]{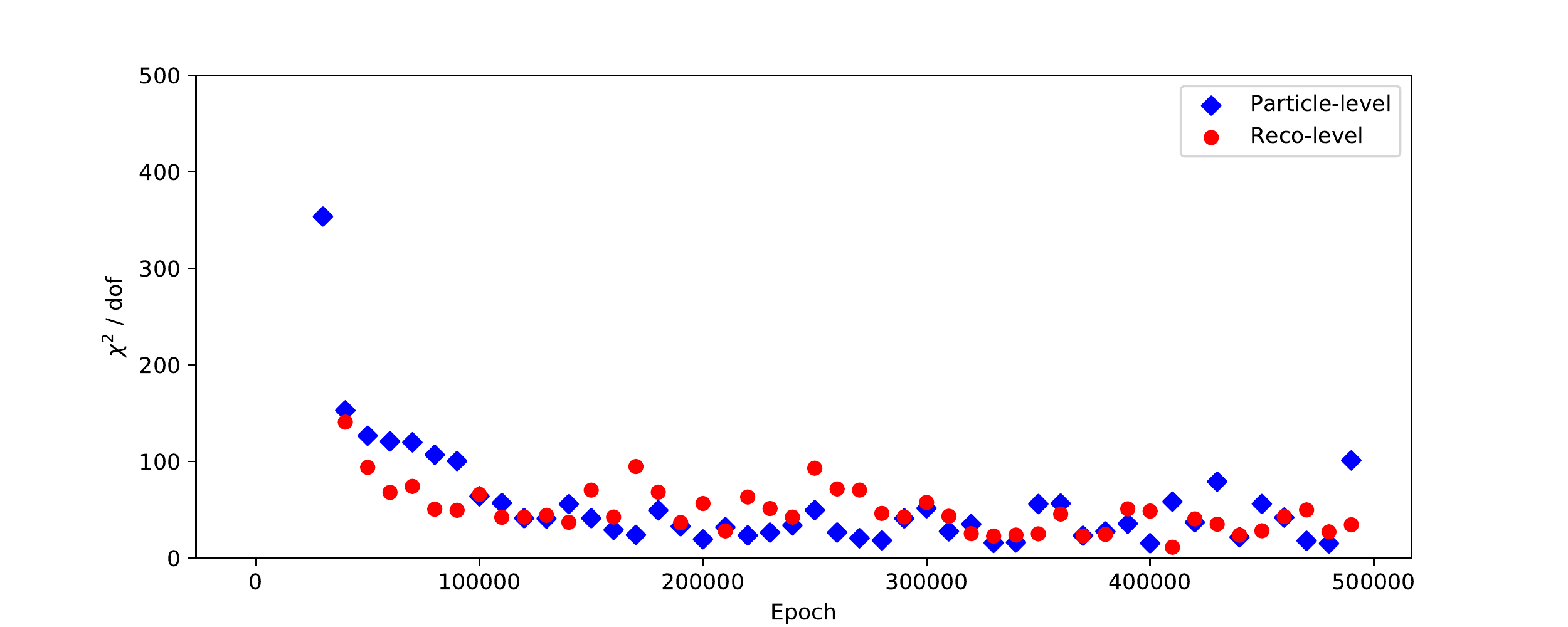}
\caption{\small{Evolution of $\chi^2$ as a function of training epoch at particle- and reco-level.}}
\label{fig:training:chi2}
\end{figure}

The network was then trained for 500,000 iterations with mini-batches of 128 events each, drawn from the original distribution and from the noise-generated fakes. It took about five hours to complete the training on a GPU NVIDIA Quadro P6000. For each iteration, we first trained the discriminator to distinguish between real and fake events. Then, the discriminator weights are fixed and the generator is trained. Figs. \ref{fig:training:ptcl} and \ref{fig:training:reco} show the discriminator and the generator loss as a function of the training epoch at particle and reco level respectively. The stationary state between the generator and the discriminator, also known as Nash equilibrium, is reached after a few thousands epochs. However, the agreement between the MC- and GAN-generated distributions improves in terms of $\chi^2$ as the number of iterations increases, and levels out after about 100,000 epochs, as shown in Fig. \ref{fig:training:chi2}. This is true for both particle and reco level, and can be easily understood as a consequence of the stabilization of the generator losses at around the same epoch.

\FloatBarrier

\section{Event Generation and Final Results}\label{sec:results}

During the training, the weights of the generator model are saved into a file every 5000 epochs and used subsequently to generate an arbitrary number of events. It takes about 80 seconds to generate 1 million events on a GPU NVIDIA Quadro P6000. After the generation, events are filtered by applying the same kinematic cuts we applied to the real MC events, \ie both jets with $p_T >$ 250 GeV, ordered by decreasing $p_T$. Approximately 90\% fulfill these requirements and are used to fill the histograms. 

\begin{figure}[htbp]
\centering
\includegraphics[scale=0.30]{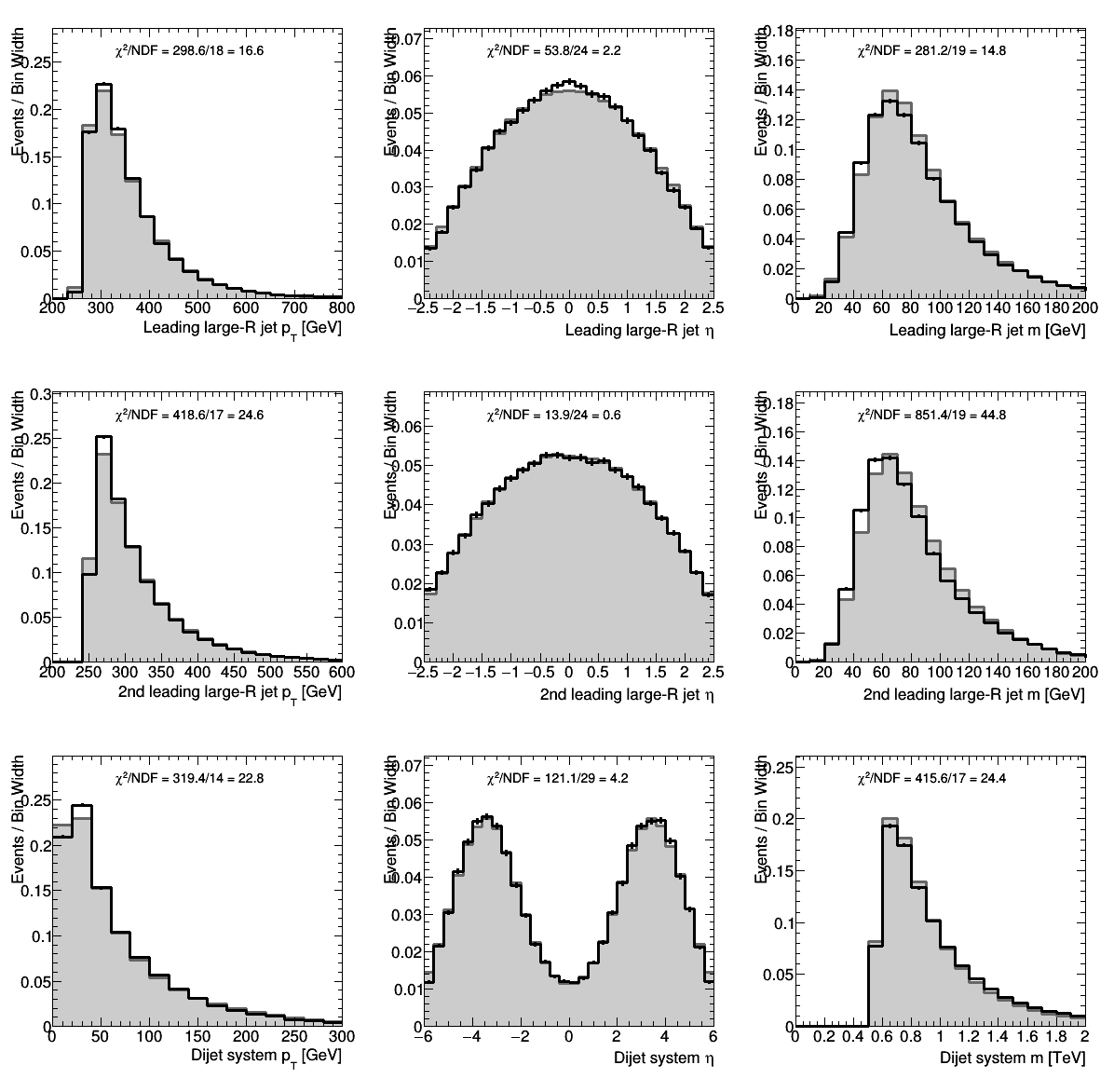}
\caption{\small{Comparison of kinematic observables with respect to particle-level (\MadGraph{} + \Pythia) Monte Carlo simulation. The gray area represents the MC prediction, and the black line indicates the GAN output.}}
\label{fig:observables:ptcl}
\end{figure}

\begin{figure}[htbp]
\centering
\includegraphics[scale=0.30]{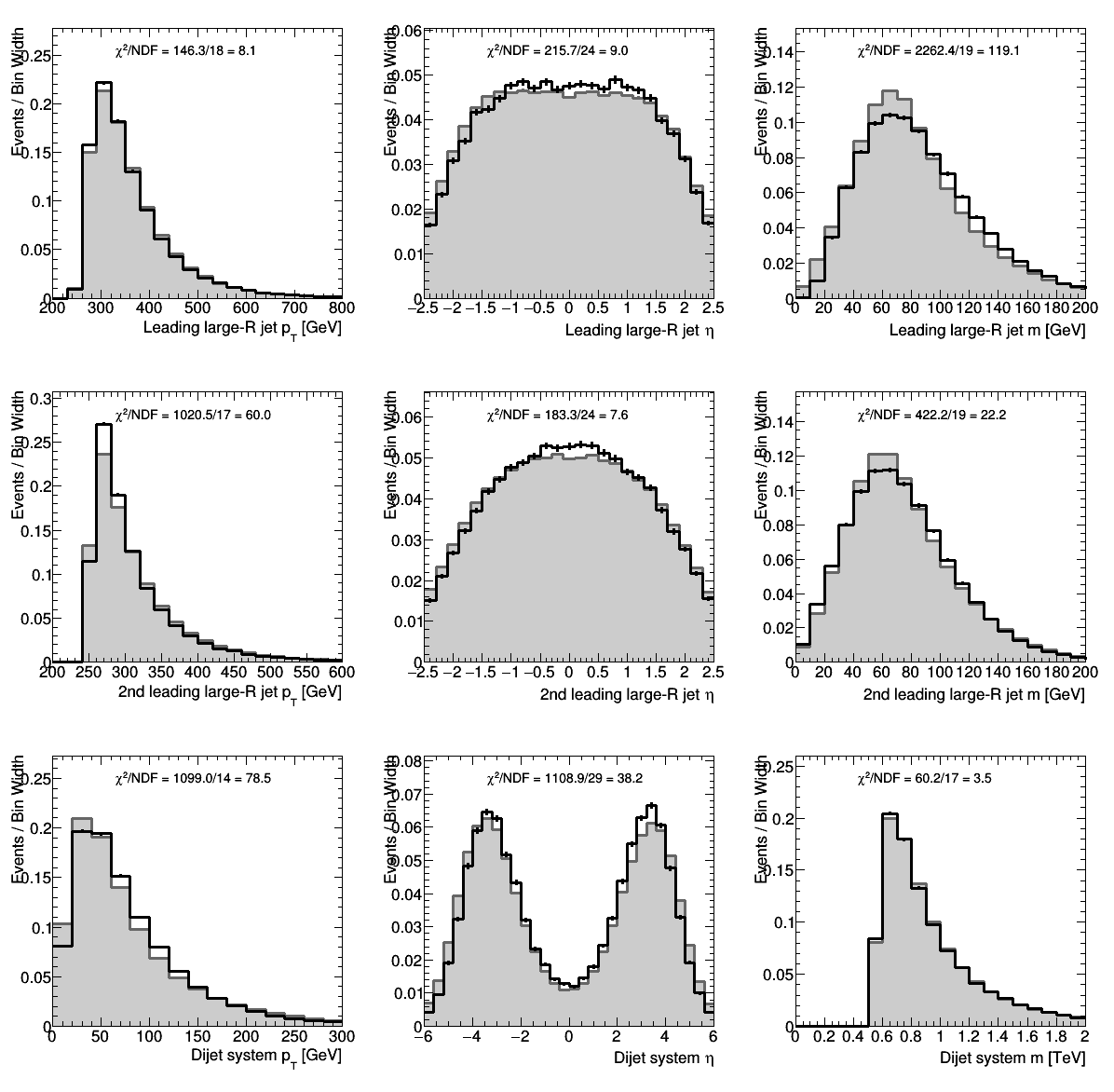}
\caption{\small{Comparison of kinematic observables with respect to reco-level (\MadGraph{}+\Pythia + \Delphes) Monte Carlo simulation. The gray area represents the MC prediction, and the black line indicates the GAN output.}}
\label{fig:observables:reco}
\end{figure}

Figs. \ref{fig:observables:ptcl} and \ref{fig:observables:reco}  show the comparison of the two leading jets and dijet system kinematics at particle- and reco-level respectively, as they appear at the iteration that yields the best agreement in terms of overall $\chi^2$ over degrees of freedom. 
Overall, the level of agreement is satisfactory over a large range of the kinematic regime.

\FloatBarrier

\paragraph{}
We further investigated the agreement in regions of the phase space with low cross-section, in particular where the dijet invariant mass is in the multi-TeV regime. This kinematic region is of particularly interest for searches of physics beyond the SM. A very common approach is to fit the MC sample with the following four-parameters ($4p$) analytic function:
\begin{equation}
    f(x) = \frac{ p_0 (1-x)^{p_1}}{x^{(p_2+p_3\log x)} }
\end{equation}

where $x = m_{jj}/\sqrt{s}$ and $p_0$, $p_1$, $p_2$, $p_3$ are the free parameters of the fit. Such function is motivated by the structure of parton distribution functions and has been widely used by Tevatron and LHC experiments \cite{DijetFitFunctions}.  

We trained the GAN using only a small fraction of the available events, about 15\% corresponding to about 150,000 events with $m_{jj}>$1.5 TeV. Then, we used the trained model to generate a sample of about 1 million events, a number comparable to the number of events in the \MadGraph{}+\Pythia{} training sample. 

\begin{figure}[htbp]
\centering
\includegraphics[scale=0.20]{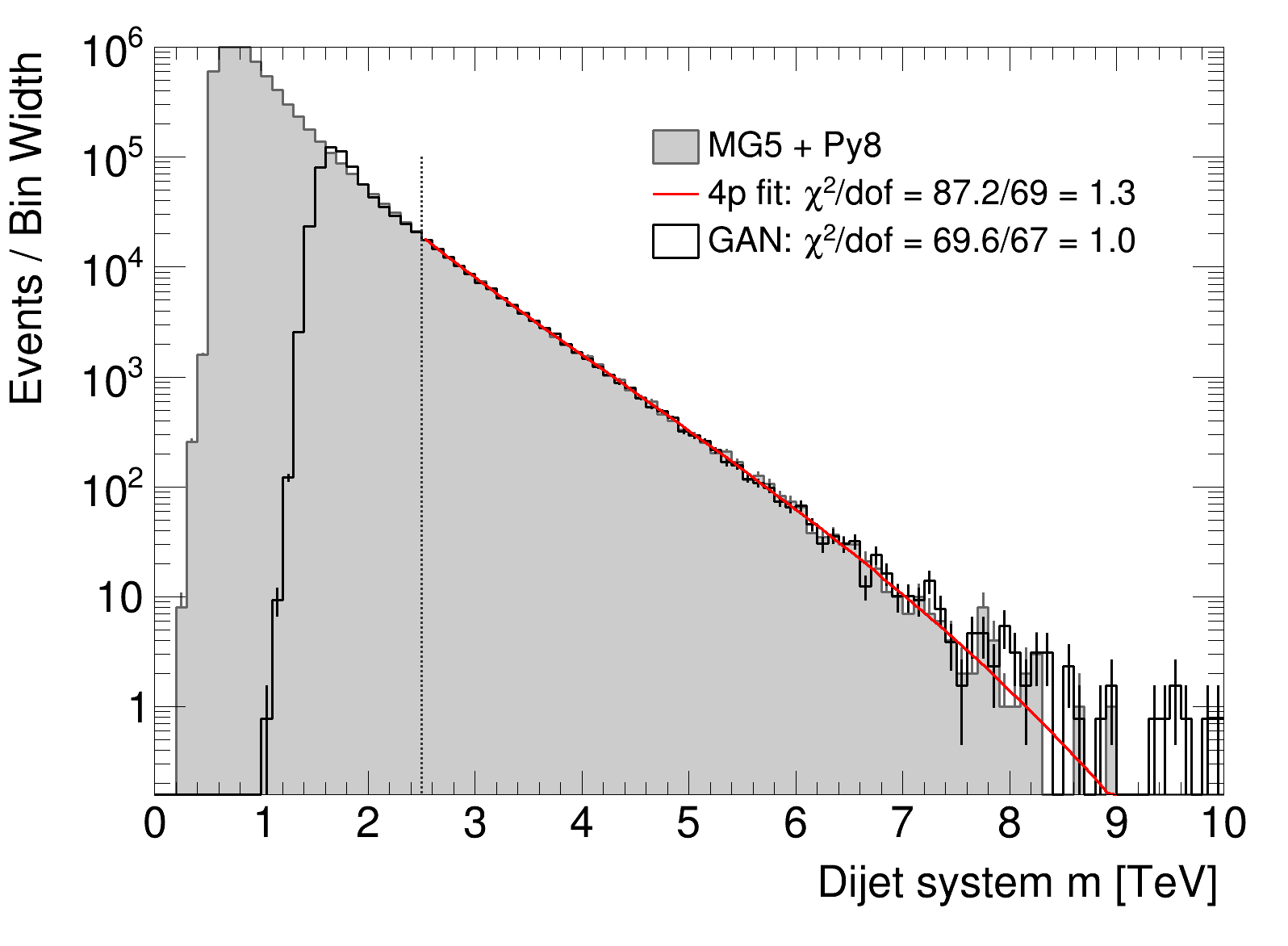}
\caption{\small{Comparison between a (\MadGraph{} + \Pythia) Monte Carlo simulation sample and the GAN extrapolation to high dijet invariant mass. The gray area represents the MC prediction, the black line indicates the GAN output, and the red line is the fitted four-parameters analytic function.}}
\label{fig:extrapolation:ptcl}
\end{figure}

As shown in Fig.\ref{fig:extrapolation:ptcl}, limiting the fit in the region between 2.5 and 10 TeV, the $4p$ analytic function can predict the shape of the MC distribution with a $\chi^2/$NDF = 1.3. In the same kinematic region, the sample generated with the GAN shows and agreement with similar $\chi^2/$NDF. Besides the agreement in a single variable, one has to take in mind that the $4p$ fit does not allow the user to generate event, but only to make an estimate of the background due to multijet production in that particular kinematic region and only for that specific observable. Therefore, events produced with our GAN can significantly expand the techniques used by analysis teams in determining their background.

\FloatBarrier

\section{Possible extensions of the method}
The baseline architecture described in Sec. \ref{sec:architecture} can be modified for more advanced purposes that go beyond the scope of this initial work, but are relevant for practical usage in collider experiments. Both extensions were implemented and produced results compatible with those presented above. The corresponding code is available in the repository.

\subsection{Arbitrary number of input variables}
In future application of the method to processes whose final states involve more than two particles, it would be desirable to have a more generic handling of the input variables. A common way to achieve this is to apply a Principal Component Analysis (PCA) to the input vector, a procedure that is often referred to as {\sl whitening} in Deep Learning  \cite{whitening}. The purpose is to reduce the dimensionality of the input vector, while at the same time retaining most of the  information needed to extract useful features. PCA can be generalized to a non-linear transformation by using an unsupervised Deep Learning technique called autoencoder \cite{autoencoder_gan}. A deep network (encoder) transforms the input in the visible representation $x_{vis}\in\mathbb{R}^N$ to a corresponding latent representation $x_{lat}\in\mathbb{R}^M$ with lower dimensionality, \ie $M<N$. Subsequently, the latent vector is fed into another network (decoder) that transforms back to the visible space. The complete chain of transformation is thus $\mathbb{R}^N\rightarrow\mathbb{R}^M\rightarrow \mathbb{R}^N$. For the purpose of this application, the generator and discriminator networks are trained using events transformed into the latent space representation ($\mathbb{R}^M$) using the encoder network. After the GAN is trained, the generator is used to create events as described above. However, before filling the histograms, the generated events are transformed back to the physical representation ($\mathbb{R}^N$) using the decoder network.

\subsection{Conditioning on external variable}
It is a common problem in searches for physics beyond the SM and precision measurements that regions of the phase space with large invariant mass or transverse momentum (typically in the TeV regime) have very low cross-sections and are hence under-represented in the simulated samples, yielding a large statistical uncertainty that may limit the sensitivity. Two classical solutions are either to generate very large samples and discard uninteresting events, or to bias the event generation by over-sampling certain regions of the phase-space. The first, brute-force approach, usually requires an unusually large CPU and disk space usage, while the second method requires events to be weighted, which can yield large fluctuations and uncertainties introduced by the event selection. A third possibility is to create so-called {\sl sliced} samples, \ie the event generation is split into a number of sub-samples in which the value of a certain variable that controls the event kinematics (such as the invariant mass or the transverse momentum of the dijet system) is limited to a certain range. The sub-samples are then added together to obtain an inclusive sample with comparable statistical uncertainty all over the range of certain variables of interest (\eg the transverse momentum of the leading jet).
In this regard, Generative-Adversarial Networks offer a similar possibility to bias the event generation without introducing event weights by the means of the so-called {\sl variable conditioning}~\cite{conditional_gan}. This is achieved by adding an auxiliary parameter to the input that controls the way noise is transformed into physical events. Examples of such conditioning parameter, as in the case of sliced MC samples, are the invariant mass or the transverse momentum of the dijet system, or the average number of pile-up interactions per bunch crossing $\left < \mu \right >$. In the context of calorimeter simulations, the GAN can be conditioned on the energy of the incident particle to generate showers corresponding to a specific energy\cite{GAN7}.
The number of conditioning parameters does not have to be limited to just one: for example, as in the case of the phenomenological supersymmetric standard model (pMSSM) \cite{BSM:SUSY}, the generator network may be conditioned to create events for any given pair of $(m_{\tilde{g}},m_{\tilde{\chi}^0})$ masses. This could be further extended to include more supersymmetric parameters.

\FloatBarrier

\section{Conclusions and Outlook}
The Generative-Adversarial Network presented in this paper provides an attractive solution to reduce the usage of CPU and possibly disk space to generate and simulate events at the LHC experiments. While still in its infancy, this method provides a unique opportunity to improve the quality of the MC used by the LHC collaborations as they will be able to use generators that are currently too time consuming to use. In the future, it should be possible to generalize this approach to more complicated processes such as top-quark pair or vector boson production in association with jets; the best MC predictions of these processes are also limited by high CPU requirements. 
Our results comparing simulated events show that our  GAN can reproduce simulated events with high accuracy. This proof-of-concept shows the potential of these tools to provide an efficient solution to the large number of simulated events required by the ambitious physics programme of the LHC experiments. 
Future work will also focus on more advanced methods to further stabilize the training and avoid model collapse, while still being able to fit the relevant kinematic distributions in regions of the phase-space with low cross-sections. 

\section*{Acknowledgements}
We acknowledge the support of the Natural Sciences and Engineering Research Council of Canada (NSERC) and of the Science, Technology and Facility Council (STFC). This project has received funding from the European Union Horizon 2020 research and innovation programme under grant agreement No 765710. We received the donation of two P6000 GPU cards (one per group) in support of our work by the NVIDIA Corporation GPU grant programme. We would like to thank Benjamin Nachman for the useful discussions.

\section*{Appendix}
\begin{figure}[htbp]
\centering
\includegraphics[scale=0.30]{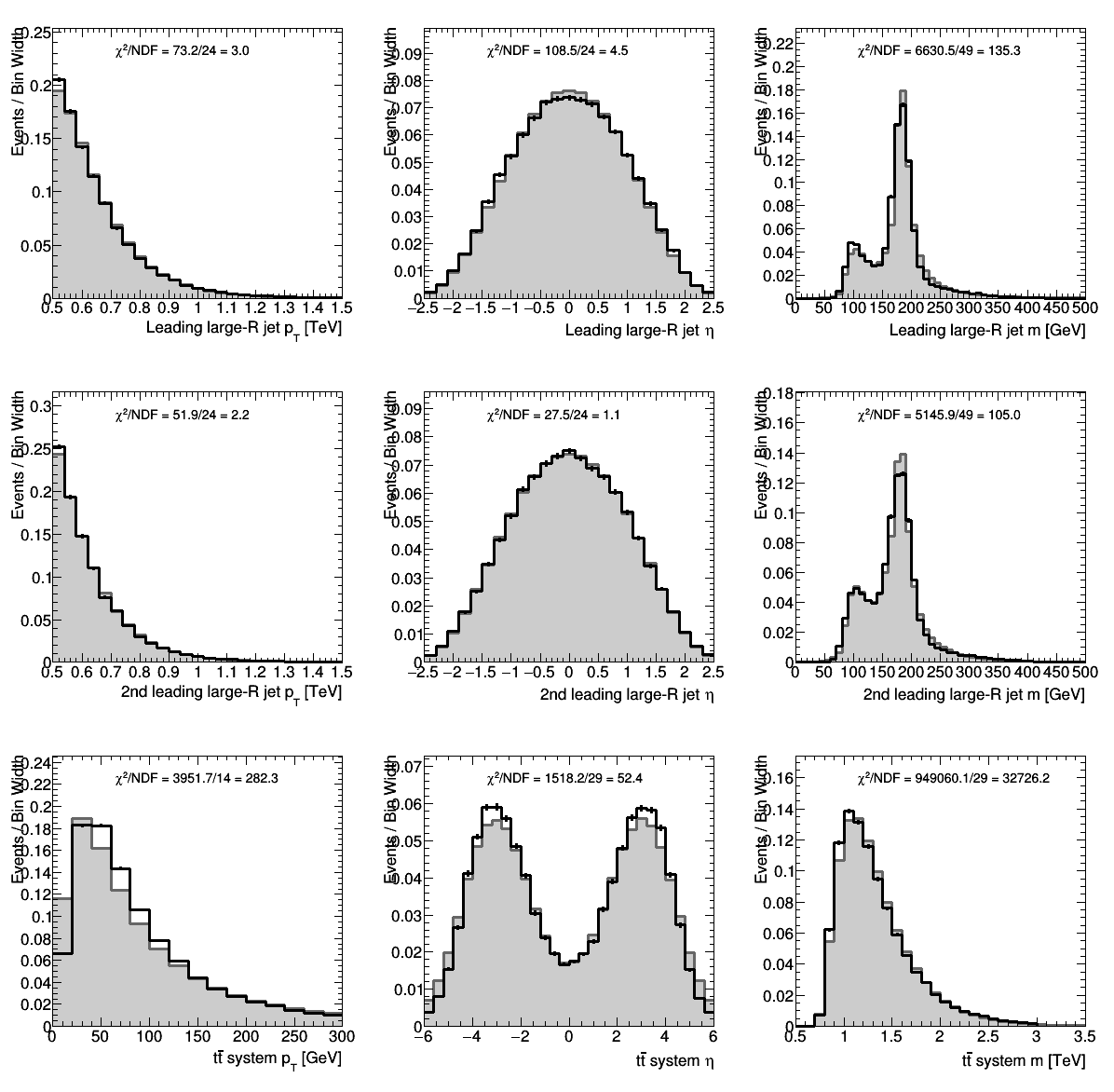}
\caption{\small{Comparison of kinematic observables for the all-hadronic $t\bar{t}$ production with respect to particle-level (\MadGraph{}+\Pythia) Monte Carlo simulation. The gray area represents the MC prediction, and the black line indicates the GAN output.}}
\label{fig:observables:ttbar}
\end{figure}

To complement the investigations described above, we also trained our GAN on a sample of top-quark pairs decaying in the all-hadronic channel, \ie $t\bar{t}\rightarrow WbW\bar{b} \rightarrow bq\bar{q}^\prime \bar{b}q\bar{q}^\prime $. To ensure that the decay products of the top quarks are collimated due to Lorentz boost in a radius $R\leq 1.0$ with respect to the jet axis, a cut on the scalar sum of the transverse momenta of the outgoing partons $H_T > 700$ GeV was applied. Also, both jets at particle level are required to have a transverse momentum $p_T>$ 350 GeV and mass $<500$ GeV. In this region of the phase space, the jet mass is expected to have a peak around the top mass, which is set to $172.5$ GeV in the MC simulation. Such configuration is known as "fully contained" top quarks. However, in some cases the $b$-quarks are produced at an angle such that only the $W$ boson is actually found within $\Delta R<$1.0 from the jet axis. In this case, a secondary peak appears around the $W$ boson mass, set to $80.4$ GeV in the MC simulation. The total jet mass distribution is then bi-modal. 
Apart from these differences, overall the events at particle level look very similar to QCD dijet ones, hence our GAN should be able to deal with this alternative physics process. It is worth stressing the fact that the information about the top quark and $W$ boson masses are not fed into the GAN, but have to be inferred by the network during the training. As can be seen from Fig. \ref{fig:observables:ttbar}, the agreement between the \MadGraph{}+\Pythia MC and the GAN output is in fact satisfactory.   

\begin{thebibliography}{99}

\bibitem{GAN} Generative Adversarial Networks, I. J. Goodfellow, et al., Proceedings of NIPS 2014, arXiv:1406.2661

\bibitem{GAN:images} A Style-Based Generator Architecture for Generative Adversarial Networks, T. Karras et al., arXiv:1812.04948

\bibitem{GAN:music} C-RNN-GAN: Continuous recurrent neural networks with adversarial training, O. Mogren, arXiv:1611.09904, Accepted to Constructive Machine Learning Workshop (CML) at NIPS 2016

\bibitem{GAN1} Learning Particle Physics by Example: Location-Aware Generative Adversarial Networks for Physics Synthesis, L. de Oliveira, M. Paganini, B. Nachman, Comput Softw Big Sci (2017) 1:4

\bibitem{GAN2} Accelerating Science with Generative Adversarial Networks: An Application to 3D Particle Showers in Multi-Layer Calorimeters, Michela Paganini, Luke de Oliveira, and Benjamin Nachman, Phys. Rev. Lett. 120, 042003 (2018)

\bibitem{GAN3} CaloGAN: Simulating 3D high energy particle showers in multilayer electromagnetic calorimeters with generative adversarial networks, Michela Paganini, Luke de Oliveira, and Benjamin Nachman, Phys. Rev. D 97, 014021

\bibitem{GAN4} Generating and refining particle detector simulations using the Wasserstein distance in adversarial networks, Martin Erdmann, Lukas Geiger, Jonas Glombitza, David Schmidt, arXiv:1802.03325

\bibitem{GAN5} Precise simulation of electromagnetic calorimeter showers using a Wasserstein Generative Adversarial Network, Martin Erdmann, Jonas Glombitza, Thorben Quast, arXiv:1807.01954

\bibitem{GAN6} Three dimensional Generative Adversarial Networks for fast simulation, Federico Carminati et al., J.Phys.Conf.Ser. 1085 (2018) no.3, 032016

\bibitem{GAN7} Deep generative models for fast shower simulation in ATLAS, ATL-SOFT-PUB-2018-001

\bibitem{GAN8} LHC analysis-specific datasets with Generative Adversarial Networks, Bobak Hashemi et al., arXiv:1901.05282

\bibitem{VAE} Auto-Encoding Variational Bayes, D. P. Kingma and M. Welling, arXiv:1312.6114

\bibitem{BetaVAE} Event Generation and Statistical Sampling with Deep Generative Models, S. Otten et al., arXiv:1901.00875

\bibitem{DCGAN} Unsupervised Representation Learning with Deep Convolutional Generative Adversarial Networks, Alec Radford, Luke Metz, Soumith Chintala, arXiv:1511.06434


\bibitem{JetTagging2} Jet-Images -- Deep Learning Edition, L. de Oliveira et al., JHEP 07 (2016) 069



\bibitem{SM1} Measurements of $t\bar{t}$ differential cross-sections of highly boosted top quarks decaying to all-hadronic final states in pp collisions at $\sqrt{s}$ = 13 TeV using the ATLAS detector, The ATLAS Collaboration, Phys. Rev. D 98, 012003 (2018)

\bibitem{SM3} Top-quark mass measurement in the all-hadronic $t\bar{t}$ decay channel at $\sqrt{s}$ = 8 TeV with the ATLAS detector, The ATLAS Collaboration, JHEP09 (2017) 118

\bibitem{SM2} Measurement of the top-quark mass in all-jets tt events in pp collisions at $\sqrt{s}$ = 7 TeV, The CMS Collaboration, Eur. Phys. J. C 74 (2014) 2758

\bibitem{BSM1} Search for new phenomena in dijet events using 37 fb$^{−1}$ of pp collision data collected at $\sqrt{s}$ = 13 TeV with the ATLAS detector, The ATLAS Collaboration, Phys. Rev. D 96, 052004 (2017)

\bibitem{BSM2} Search for dijet resonances in proton--proton collisions at $\sqrt{s}$ = 13 TeV and constraints on dark matter and other models, The CMS Collaboration, Physics Letters B 769, 520-542 (2017)

\bibitem{BSM3} Search for supersymmetry in the all-hadronic final state using top quark tagging in pp collisions at $\sqrt{s}$ = 13 TeV, The CMS Collaboration, Phys. Rev. D 96, 012004 (2017)

\bibitem{BSM4} Search for $W^\prime \to tb$ decays in the hadronic final state using pp collisions at $\sqrt{s}$ = 13 TeV with the ATLAS detector, The ATLAS Collaboration, Phys.Lett. B781 (2018) 327-348

\bibitem{BSM5} Search for new resonances decaying into boosted W, Z and H bosons at CMS, M. Krohn and C. Vernieri, FERMILAB-CONF-17-429-PPD, arXiv:1710.02217

\bibitem{ATLAS} The ATLAS experiment at the CERN Large Hadron Collider, ATLAS Collaboration, J Instrum, 2008, 3: S08003

\bibitem{CMS} The CMS experiment at the CERN LHC, The CMS Collaboration, J Instrum, 2008, 3: S08004

\bibitem{MadGraph} MadGraph 5 : Going Beyond, J. Allwall et al., JHEP 06 (2011) 128

\bibitem{POWHEG-BOX} A general framework for implementing NLO calculations in shower Monte Carlo programs: the POWHEG BOX, S. Alioli, P. Nason, C. Olearic, E. Re, JHEP 06 (2010) 043

\bibitem{MCatNLO} Matching NLO QCD computations and parton shower simulations, S. Frixione, B.R. Webber, JHEP 06 (2002) 029

\bibitem{Pythia8} A Brief Introduction to PYTHIA 8.1, T. Sj\"ostrand, S. Mrenna and P. Z. Skands, Comput. Phys. Commun. 178 (2008) 852

\bibitem{Herwig7} Herwig 7.0/Herwig++ 3.0 release note, J. Bellm et al., Eur. Phys. J. C 76 (2016) 196

\bibitem{Sherpa} Event generation with Sherpa 1.1, T. Gleisberg et al., JHEP 02 (2009) 007

\bibitem{jetography} Towards Jetography, G. Salam, Eur. Phys. J. C67 (2010) 637

\bibitem{Geant4} GEANT4 Collaboration, Nuclear Instruments and Methods in Physics Research A 506, 250 (2003).

\bibitem{hsf-mc} A Roadmap for
HEP Software and Computing R\&D for the 2020s, The HEP software foundation, arXiv:1712.06982v5

\bibitem{fxfx} Merging meets matching in MC@NLO, R. Frederix and S. Frixione, JHEP12(2012)061

\bibitem{mepsnlo} Uncertainties in MEPS@NLO calculations of $h$+jets, S. H\"{o}che, F. Krauss and M. Sch\"{o}nherr, Phys. Rev. D 90, 014012

\bibitem{unlops} Merging Multi-leg NLO Matrix Elements with Parton Showers, L. Lonnblad and S. Prestel, JHEP03(2013)166

\bibitem{Delphes} DELPHES 3: a modular framework for fast simulation of a generic collider experiment, J. de Favereau et al., J. High Energ. Phys. (2014) 2014: 57

\bibitem{antikt} The anti-$k_t$ jet clustering algorithm, Matteo Cacciari, Gavin P. Salam and Gregory Soyez, Journal of High Energy Physics, Volume 2008, JHEP04(2008)

\bibitem{fastjet} FastJet: a code for fast $k_t$ clustering, and more, M. Cacciari, arxiv:hep-ph/0607071

\bibitem{LeakyReLU} Empirical Evaluation of Rectified Activations in Convolutional Network, Bing Xu et al., arXiv:1505.00853

\bibitem{Keras} F. Chollet, (2015) Keras, GitHub. \url{https://github.com/fchollet/keras}

\bibitem{tensorflow} TensorFlow: Large-Scale Machine Learning on Heterogeneous Systems, M. Abadi et al. (2015), \url{https://www.tensorflow.org}

\bibitem{scikit-learn} Scikit-learn: Machine Learning in {P}ython, F. Pedregosa et al., Journal of Machine Learning Research vol. 12 (2011)

\bibitem{pandas} Data Structures for Statistical Computing in Python, Wes McKinney, Proceedings of the 9th Python in Science Conference (2010)

\bibitem{adam} Adam: A Method for Stochastic Optimization, D. P. Kingma and J. Ba, arXiv:1412.6980, conference paper at the 3rd International Conference for Learning Representations, San Diego, 2015

\bibitem{whitening} Optimal Whitening and Decorrelation, A. Kessy et al., The American Statistician, 72: 309-314 (2018)

\bibitem{autoencoder_gan} Generating Faces with Torch, A. Boesen et al., 2015, \url{torch.ch/blog/2015/11/13/gan.html}

\bibitem{conditional_gan} Conditional Generative Adversarial Nets, M. Mirza and S. Osindero, arXiv:1411.1784

\bibitem{BSM:SUSY} The BSM-AI project: SUSY-AI - Generalizing LHC limits on Supersymmetry with Machine Learning, S. Caron et al., Eur. Phys. J. C (2017) 77: 257

\bibitem{DijetFitFunctions} Searches for Dijet Resonances at Hadron Colliders, R. H. Harris and K. Kousouris, Int. J. Mod. Phys. A26:5005-5055, 2011

\end{thebibliography}
\end{document}